\documentstyle[12pt]{article}

\def\BR{{\mathchoice
	{\setbox0=\hbox{$\displaystyle\rm  R$}
  \hbox{\hbox to0pt{\kern-0.2\wd0\rm 
I\hss}\box0}} {\setbox0=\hbox{$\textstyle\rm 
R$}\hbox{\hbox to0pt{\kern-0.2\wd0$\textstyle\rm
I$\hss}\box0}} {\setbox0=\hbox{$\scriptstyle\rm 
R$}\hbox{\hbox to0pt{\kern-0.2\wd0$\scriptstyle\rm
I$\hss}\box0}} {\setbox0=\hbox{$\scriptscriptstyle\rm 
R$}\hbox{\hbox to0pt{\kern-0.15\wd0$\scriptscriptstyle\rm
I$\hss}\box0}}}} 

\newcommand {\be}{\begin{displaymath}}
\newcommand {\ee}{\end{displaymath}}
\newcommand {\ben}{\begin{equation}}
\newcommand {\een}{\end{equation}}
\newcommand {\bea}{\begin{eqnarray*}}
\newcommand {\eea}{\end{eqnarray*}}
\newcommand {\bean}{\begin{eqnarray}}
\newcommand {\eean}{\end{eqnarray}}

\newcommand {\VL}{de Sousa Vieira and Lichtenberg}

\begin{document}
\title{Linear and optimal nonlinear control of one-dimensional maps}
\author{
Justin McGuire, Murray T. Batchelor, Brian Davies\\
Department of Mathematics\\
School of Mathemtical Sciences\\
Australian National University\\
Canberra ACT 0200, Australia\\}
\date{March 21, 1997}
\maketitle

\begin{abstract}
We investigate the effects of linear and optimal nonlinear
control in the simple case of one-dimensional unimodal maps.
We show that linear feedback relates unimodal maps to invertible
``Henon-like'' maps.
This observation should be useful in relating the considerable bodies of
knowledge which exist for the two types of systems.
In the case of the optimal nonlinear feedback scheme of \VL\ we
investigate the relationship between controlled and uncontrolled maps,
particularly the preservation of the period doubling route to chaos.
\end{abstract}

\vspace{1cm}
\noindent
PACS: 05.45.+b

\section{Introduction}

In recent years there has been much attention given to theoretical
and experimental methods for stabilizing unstable periodic orbits
(UPOs) of chaotic systems (see, e.g., 
[1-11] and refs
therein). The methods typically operate by feedback perturbation applied
either to an available system parameter \cite{OGY,BDG} or to a state
variable
\cite{Pyr92,SVL}. 
Pyragas \cite{Pyr92} suggested delayed feedback
control, which incorporates memory into the system.
This idea was extended by Socolar, Sukow and Gauthier \cite{SSG};
and \VL\  \cite{SVL} (amongst others).  

These methods have been the subject of much investigation, both
theoretical and experimental, and have proved to be very versatile.
An important practical consideration is that 
feedback can be applied quite simply to physical systems.
The size of the perturbation is determined by
a comparison of the current state of the system with the state of
the system at $\tau$ in the past, where
$\tau$ is the period of the desired UPO,
although the details vary according to the method. 

This paper deals with the effect of two different control methods. 
The first adds a linear perturbation to the system, so
that the system $x_{n+1}=f(x_n)$ takes the form
$x_{n+1}=f(x_n)+k(x_n-x_{n-1})$. 
Here $k\in[0,1)$ is a parameter determining how strongly the system
is controlled. 
This is the return map version of the feedback technique suggested
by Pyragas \cite{Pyr92}, which we will refer to as linear control.
The second method adds a nonlinear perturbation, which gives the
system the form $x_{n+1}=f(x_n)+k(x_n-f(x_n))$. 
This is the optimal control method suggested by \VL\ \cite{SVL}.

Both of these methods are capable of successfully stabilizing the fixed
points of quite general systems. 
We show that for some of the unimodal maps on which they have
been tested previously they have the effect of producing
systems which are already well-known.
This observation links these particular controlled systems with
others for which there is a body of theoretical understanding.

\section{Unimodal Maps with Linear Control}
A linearly controlled one-dimensional map can be written as a function 
from $\BR^2$ to $\BR^2$ as follows
\bean
x_{n+1}&=&f(x) + k(x_n-y_n)\label{cm1}\\
y_{n+1}&=&x_n\label{cm2}
\eean
where $y$ is used to keep track of the previous iteration.
This is similar to what we shall call a {\em generalized Henon map}
\bean 
X_{n+1}&=&H(X_n)+Y_n\label{h1}\\
Y_{n+1}&=&BX_n\label{h2}\\
H(0)&=&1\hskip1cm\hbox{(local maximum)}\label{h3}
\eean
For $B\ne0$, this generalized Henon map is invertible, with Jacobian $B$.
Thus, those properties of the Henon map which depend on this will
generalize.

We shall show that the two systems are equivalent under a simple linear
transformation.  
The most general form giving equivalence of (\ref{cm2}) and (\ref{h2}) is 
\bea
x&=&\alpha BX+\beta\\
y&=&\alpha Y+\beta
\eea
with 
\be
B=-k
\ee
from which 
\be
H(X)=-\frac{\beta}{\alpha B}-BX+\frac{f(\alpha BX + \beta)}{\alpha B}.
\ee
The values of $\alpha$ and $\beta$ are determined by applying the
conditions (\ref{h3}).

In the case that $f(x)$ is differentiable, these conditions become
$H(0)=1$, $H'(0)=0$.  
For example, from the logistic map
\ben
f(x)=rx(1-x)\label{lmap}
\een
we get the standard Henon map \cite{Henon}, for which $H(X)=1-AX^2$, with
\bea
\alpha&=& \frac{(k+r)(2+k-r)}{4kr}\\
\beta &=& \frac{k+r}{2r}\\
A&=&\frac{(k+r)(r-2-k)}{4}
\eea
so the two are in fact equivalent.

The generalized tent map
\ben
f(x)=\left\{
\begin{array}{lll}
t\displaystyle{\frac{x}{\raise4pt\hbox{$s$}}}&\hspace{1cm}&(0<x<s)\\
t\displaystyle{\frac{(1-x)}{(1-s)}}&&(s<x<1)
\end{array}\right. \label{tent defn}
\een
is an example where the function $f$ is not differentiable.
Applying the conditions (\ref{h3}) gives
\be
\alpha=\frac{s-t}{k}
\hspace{2cm}
\beta=s
\ee
so we get a Lozi-style map
\bea
H(X)=\left\{
\begin{array}{lll}
1+\Big(k+\displaystyle{\frac{t}{s}}\Big)X&\hspace{1cm}&(X<0)\\
1+\Big(k+\displaystyle{\frac{t}{s-1}}\Big)X&&(X>0)
\end{array}
\right.
\eea
The regular Lozi map has the form \cite{Lozi}
\be
H(X)=1-A|X|
\ee
Which requires $k+t/s=-k-t/(s-1)$.
Thus for a given $A$ and $B$ we have
\be
s=\frac{A-B}{2A}
\hspace{2cm}
t=\frac{A^2-B^2}{2A}
\ee
The interesting point to note from all this is that the often-studied
Henon and Lozi maps are in fact standard one-dimensional maps 
{\em destabilised} by linear feedback ($k<0$). 

\section{Unimodal Maps with Optimal Control}
In their paper on nonlinear feedback control \cite{SVL} \VL\  suggested a 
control method which combines nonlinear feedback with memory \cite{SSG}, of
the form
\bea
x_{n+1}&=&f(x_n)+\epsilon_n\\
\epsilon_{n+1}&=&-k[f(x_{n+1})-f(x_n)]+\ell\epsilon_n.
\eea 
This two-dimensional system works by adding a perturbation based not
only on the state of the system at one point in the past, but also on
previous perturbations. Here $\ell\in [0,1]$ is an additional parameter
which determines the weighting given to the previous perturbation.

They  looked for a way to make this system superstable at its fixed point. 
When $\ell\not=0$ this occurs only if $\ell=k$, in which case the
perturbation $\epsilon_n$ becomes entirely dependant on $x_n$ and the
system becomes one-dimensional. 
They referred to this as optimal control.\footnote{Note that here the
word does not have its more usual meaning from control theory.} 
For a one dimensional map $x_{n+1}=f(x_n)$ this optimal control scheme is
given by 
\be x_{n+1}=(1-k)f(x_n)+kx_n.
\ee
This method has the advantage that it is very simple and, for the correct
choice of $k$, can make the fixed point not only stable but superstable,
regardless of how negative $f'$ is at the fixed point.
They applied it to the logistic map, calculating the basins of
attraction, upper limit of stability and Lyapunov exponents for
particular $k$ and $r$ of this new controlled system. 
Here we focus on the fact that this form of control is a reduction of
dimension.

The optimally controlled logistic map (\ref{lmap}) is given by
\ben
x_{n+1}=(1-k)rx_n(1-x_n)+kx_n. \label{optdefn}
\een
If we make the transformation $x=\alpha X$, where 
\be
\alpha=\frac{r(1-k)+k}{r(1-k)}
\ee
it becomes
\ben
X_{n+1}=RX_n(1-X_n) \label{Rmap}
\een
where $R=r(1-k)+k$.
With this transformation, we can determine all the properties of the
controlled logistic map from those of the uncontrolled map.

This observation is not limited to the logistic map. 
In a similar way to the above treatment of linear control, we investigate
the circumstances under which a unimodal map with optimal control can be
made equivalent to an uncontrolled unimodal map 
\be X_{n+1}=F(X_n)
\ee
by the simple linear transformation
\be
x=\alpha X + \beta.
\ee
From this
\be
F(X)=\frac{(1-k)[f(\alpha X + \beta)-\beta]}{\alpha}+kX
\ee
and we determine $\alpha$ and $\beta$ using the unimodal conditions
$F(0)=F(1)=0$.  
From $F(0)=0$ we find that $\beta$ must be a fixed point of $f$; 
for simplicity, choose $\beta=0$.
From the condition $F(1)=0$ we find
\ben
\frac{f(\alpha)}{\alpha}=\frac{k}{k-1}. \label{alphaopt}
\een 
Thus, if we can find an $\alpha$ satisfying (\ref{alphaopt}), we can
make the two maps equivalent.

As a further example, for the tent map (\ref{tent defn}) we find
\be
\alpha=\frac{t(k-1)}{t(k-1)-k(s-1)}
\ee
providing that $\alpha>s$.
This gives
\be
F(X)=\left\{
\begin{array}{lll}
\displaystyle{\frac{ks+t-tk}{s}}X&\hspace{1cm}&(0<X<s/\alpha)\\
\displaystyle{\frac{(-tk+t-k+ks)}{1-s}}(1-X)&&(s/\alpha<X<1)
\end{array}\right.
\ee 
So the optimally controlled tent map it equivalent to an uncontrolled
tent map, for suitable combinations of $s$, $t$ and $k$.

\section{Schwartzian derivative}

In \cite{SVL} \VL\ found that the optimally controlled logistic map
undergoes period doubling when its fixed point became unstable. 
This is only to be expected, since the sytem is equivalent to a rescaled
logistic map. 
Here we wish to investigate the conditions under which period doubling
occurs for general unimodal maps which are being stabilised at a fixed
point using optimal control.  
Thus we must examine the relationship between the Schwartzian derivatives
$SF$ and $Sf$, where
\ben 
(Sg)(x)=\frac{g '''(x)}{g'(x)}-\frac{3}{2}
\left(\frac{g''(x)}{g'(x)}\right) ^2\label{schwartzian defn}
\een
The condition for period doubling, at a point where $F'$ decreases
through $-1$, is $SF<0$ \cite{Devaney}.
($F'=-1$ corresponds to $f'=(1+k)/(k-1)$).

Using $x=\alpha X$ 
together with (\ref{schwartzian defn}), we have
\be
(SF)(X)=\frac{\alpha^2(1-k)}{[(1-k)f'(x)+k]^2}
[(1-k)(f'(x))^2(Sf)(x)+kf'''(x)]
\ee
Therefore, for $k\in(-1,1)$, the sign of $SF$ is determined solely by
the  expression
\be
(1-k)(f'(x))^2(Sf)(x)+kf'''(x)
\ee
which is linear in $k$, having the same sign as $Sf$ at $k=0$.
The critical observation is that, for small enough $k$, the
period-doubling  property is preserved by the application of optimal
control. 
For non-zero $k$, the r\^ole of $f'''$ must be considered. 
For control (rather than destabilisation) $k\in[0,1)$. If $f'''\leq0$, we
see that period doubling is preserved for all values of $k$, whereas for
$f'''>0$ the character of the bifurcation may change at some critical
value of $k$.

The relation between $Sf$ and $SF$ in the neighhourhood of a given fixed
point is essentially a local property, and does not depend on the
requirement that the maps be unimodal.
One expects, therefore, that the preservation of the period doubling route
to chaos, when a system is subject to optimal control, should be quite
general.

\section{Conclusion}

We have investigated the effects of linear and optimal nonlinear
control in the simple case of one-dimensional unimodal maps.
Such maps are important testbeds for methods which can also be applied
in higher dimensions.
 
We have shown that the application of linear feedback turns the logistic
map into the Henon map; more generally, when applied to any unimodal
map, linear feedback results in a ``Henon-like'' map
which is invertible and for which the contraction factor is $|k|$.
For example, the linearly controlled tent map is a ``Lozi-like'' map.
It is well-known that the addition of feedback increases the
dimensionality of a system; our attention here is particularly on the
interconnections which this affords.
One might hope to obtain some precise knowledge of strange
attractors for the controlled systems in this way.
Existing investigations of strange attractors have been for what is, in effect,
a linearly destabilised map.

For nonlinear feedback, \VL\ have observed that it is possible to recover
the original dimensionality by a procedure which they named optimal control.
Having seen that the optimally controlled logistic map is precisely
equivalent to an uncontrolled logistic map with a rescaling of parameter,
we went on to investigate the relationships which exist for more general
unimodal maps, including the tent map. Of particular interest are the
conditions under which the inevitable instability of a fixed point (caused
by increasing a system parameter) leads into the period doubling route for
the controlled system, if this was the route for the original system.
In the case of many unimodal maps we have shown that the necessary
condition, that the Schwartzian derivative be negative, is preserved for
sufficiently small values of feedback parameter $|k|$ (in many cases for
all $k\in[0,1)$). 



\begin{thebibliography}{99}

\bibitem{OttBk} Coping with chaos: analysis of chaotic data and the
exploitation of chaotic systems, eds. E. Ott, T. Sauer, 
J.A. Yorke  (J. Wiley, New York, 1994).

\bibitem{OGY} E. Ott, C. Grebogi and J.A. Yorke,
Phys. Rev. Lett, 64 (1990) 1196.

\bibitem{ORGD} F.J. Romeiras, C. Grebogi, E. Ott and W.P. Dayawansa,
Physica D, 58 (1992) 165.

\bibitem{Pyr92} K. Pyragas, Phys. Lett. A, 170 (1992) 421.

\bibitem{PT} K. Pyragas and A. Tama\u{s}evi\u{c}ius, 
Phys. Lett. A, 180 (1993) 99.

\bibitem{SGOY} T. Shinbrot, C. Grebogi, E. Ott and J. A. Yorke,
Nature, 363 (1993) 411.

\bibitem{CD} G. Chen and X. Dong, Int. J. Bifurction and Chaos, 
3 (1993) 1363.

\bibitem{BDG} S. Beilawski, D. Derozier and P. Glorieux, 
Phys. Rev. E, 49 (1994) R971.

\bibitem{SSG} J.E.S. Socolar, D.W. Sukow and D.J. Gauthier, 
Phys. Rev. E, 50 (1994) 3245.

\bibitem{DYIDSG} M. Ding, W. Yang, V. In, W.L. Ditto, M.L. Spano
and B. Gluckman, Phys. Rev. E, 53 (1996) 4334.

\bibitem{SVL} M. de Sousa Vieira and A.J. Lichtenberg,
Phys. Rev. E, 54 (1996) 1200.

\bibitem{Henon} M.Henon, Comm. Math. Phys., 50 (1976) 69.

\bibitem{Lozi} R. Lozi, J. Phys. (Fr) Colloque, 39 (1978) C5-9.

\bibitem{Devaney} R.L. Devaney, 
An introduction to chaotic dynamical systems, 2nd ed.,
(Addison-Wesley, USA, 1989).

\end{thebibliography}
\end{document}